\documentclass[aps,prb,twocolumn,amsmath,amssymb]{revtex4}
\usepackage{graphicx}
\usepackage{dcolumn}

\begin{document}

\title{Nitrogen defects and ferromagnetism of Cr-doped AlN diluted magnetic semiconductor from first principles}
\author{Li-Jie Shi}\email[Present address: Institute of Semiconductors,
Chinese Academy of Sciences, 100083 Beijing, China]{}
\author{Li-Fang Zhu}\author{Yong-Hong Zhao}
\author{Bang-Gui Liu}\email[Corresponding author. Email: ]{bgliu@mail.iphy.ac.cn}
\affiliation{Institute of Physics, Chinese Academy of Sciences,
Beijing 100190, China\\
Beijing National Laboratory for Condensed Matter Physics, Beijing
100190, China}

\date{\today}

\begin{abstract}
High Curie temperature of 900 K has been reported in Cr-doped AlN
diluted magnetic semiconductors prepared by various methods, which
is exciting for spintronic applications. It is believed that N
defects play important roles in achieving the high temperature
ferromagnetism in good samples. Motivated by these experimental
advances, we use a full-potential density-functional-theory method
and supercell approach to investigate N defects and their effects
on ferromagnetism of (Al,Cr)N with N vacancies ($V_{\rm N}$). We
investigate the structural and electronic properties of the
$V_{\rm N}$, single Cr atom, Cr-Cr atom pairs, Cr-$V_{\rm N}$
pairs, and so on. In each case, the most stable structure is
obtained by comparing different atomic configurations optimized in
terms of the total energy and the force on every atom, and then is
used to calculate the defect formation energy and study the
electronic structures. Our total energy calculations show that the
nearest substitutional Cr-Cr pair with the two spins in parallel
is the most favorable and the nearest Cr-$V_{\rm N}$ pair makes a
stable complex. Our formation energies indicate that $V_{\rm N}$
regions can be formed spontaneously under N-poor condition because
the minimal $V_{\rm N}$ formation energy equals -0.23 eV or
Cr-doped regions with high enough concentrations can be formed
under N-rich condition because the Cr formation energy equals 0.04
eV, and hence real Cr-doped AlN samples are formed by forming some
Cr-doped regions and separated $V_{\rm N}$ regions and through
subsequent atomic relaxation during annealing. Both of the single
Cr atom and the N vacancy create filled electronic states in the
semiconductor gap of AlN. N-vacancies enhance the ferromagnetism
by adding 1$\mu_B$ to the Cr moment each, but reduce the
ferromagnetic exchange constants between the spins in the nearest
Cr-Cr pairs. These calculated results are in agreement with
experimental observations and facts of real Cr-doped AlN samples
and their synthesis. Our first-principles results are useful to
elucidating the mechanism for the ferromagnetism and exploring
high-performance Cr-doped AlN diluted magnetic semiconductors.
\end{abstract}

\maketitle

\section{Introduction}

The diluted magnetic semiconductor (DMS) is of great interest
because it has merits of both semiconductor and spin magnetism.
Especially the possibility of high-temperature ferromagnetism in
DMSs has inspired scientists from all the world to explore new
promising DMS materials and study their physical properties for
spintronic applications\cite{spint,sci287}. (Ga,Mn)As has been
studied substantially because of its close relation with important
GaAs semiconductor, and its ferromagnetism is believed to be
induced by hole carriers\cite{gma-dft,gma-rmp}. Generally
speaking, it is believed that at least 500 K as Curie temperature
is necessary for real spintronic devices to work at
room-temperature\cite{coey}. Curie temperature higher than 500 K
is achieved in various transition-metal doped
oxides\cite{znoa,znob,znoc,tio2a,tio2b,sno2a,sno2b,in2o3,coey1}
and nitrides\cite{gan,aln}. For Cr doped AlN, or (Al,Cr)N, high
Curie temperature of 900 K has been reported by several
groups\cite{aln900a,aln900b,aln900c}. It is shown that the high
temperature ferromagnetism in (Al,Cr)N materials is intrinsic, not
produced by secondary
phases\cite{aln,aln900a,aln900b,aln900c,aln-secod,aln-insua,aln-insub}.
Naturally, such high Curie temperatures make the materials
promising for practical spintronic applications\cite{spint}.
Although ferromagnetism of (Ga,Mn)As DMS materials is believed to
be induced by their effective carriers\cite{gma-dft,gma-rmp}, the
high temperature ferromagnetism of (Al,Cr)N materials must be
caused by different mechanism because some single-phase (Al,Cr)N
samples with high Curie temperatures are highly
insulating\cite{aln-insua,aln-insub} (the same as pure
AlN\cite{aln-insu-dft}) or in the regime of variable-range hopping
conduction\cite{alncr-vrh} (similar to (Ga,Cr)N\cite{gancr-vrh}).
The ferromagnetism should be attributed to $pd$ hybridization of
Cr $d$ states and N $p$ ones\cite{gmn-dft1,gmn-dft2,alncr-slj}. On
the other hand, defects and impurities are believed to play
important roles in DMS materials of wide semiconductor gaps
\cite{coey1,aln-insua,aln-insub,aln-othera,aln-otherb,aln-npres,two-regime,gmn-EPL}.
Nitrogen pressure during synthesis is shown to affect the
ferromagnetism in resultant
samples\cite{aln-insua,aln-insub,aln-npres}. It appears that
reduced N content enhances the ferromagnetism, being in contrast
with first-principles-calculated trend of N-vacancies in the case
of GaN-based DMSs\cite{gmn-dft3}. Therefore, it is highly
desirable to investigate effect of N content on the high
temperature ferromagnetism in (Al,Cr)N materials, and in return
its elucidation is helpful to obtain deeper insight into the
essential mechanism of the high temperature ferromagnetism.

In this paper we use a full-potential density-functional-theory
method and 3$\times$3$\times$2 supercell approach to investigate N
defects and their effects on ferromagnetism of (Al,Cr)N with N
vacancies ($V_{\rm N}$), namely Cr$_x$Al$_{1-x}$N$_{1-y}$. We
obtain the most stable structures of Al$_{36}$N$_{35}$ (one
$V_{\rm N}$ in the supercell), CrAl$_{35}$N$_{36}$ (one
substitutional Cr in the supercell), CrAl$_{35}$N$_{35}$ (one
$V_{\rm N}$ and one substitutional Cr atom in the supercell),
Cr$_2$Al$_{34}$N$_{36}$ (two substitutional Cr atoms in the
supercell), and so on by comparing their different atomic
configurations optimized in terms of the total energy and the
force on every atom, and thereby calculate their defect formation
energies. Then we investigate the structural and electronic
properties of the $V_{\rm N}$, single Cr atom, Cr-Cr pairs, and
Cr-$V_{\rm N}$ pairs. Without N vacancy, the nearest
substitutional Cr-Cr pair with their moments orienting in parallel
is the most favorable in total energy. The nearest pair of an N
vacancy and a single Cr atom forms a stable complex of Cr$^{-}$
and $V_{\rm N}^{+}$. Our calculated formation energies indicate
that $V_{\rm N}$ regions can be formed spontaneously under N-poor
condition or high enough Cr concentrations can be formed under
N-rich condition, and hence real Al$_{1-x}$Cr$_x$N$_{1-y}$ samples
are formed by forming Cr-doped regions and some separated $V_{\rm
N}$ regions and through subsequent atomic relaxation during
annealing. The doped Cr atom and the N-vacancy create filled
electronic states in the semiconductor gap of AlN. $V_{\rm N}$
enhances the ferromagnetism by giving one electron to a Cr ion to
add 1$\mu_B$ to the Cr moment, which is in agreement with
experimental observation that low N pressure during synthesis is
helpful to obtaining better ferromagnetic
Al$_{1-x}$Cr$_x$N$_{1-y}$ samples\cite{aln-insub,aln-npres}. On
the other hand, $V_{\rm N}$ reduces the ferromagnetic exchange
constants between the spins in the nearest Cr-Cr pairs, which is
consistent with experimental fact that too many N vacancies are
harmful to the ferromagnetism when the synthesis temperature is
high enough to allow doped Cr atoms to diffuse. These
first-principles results are useful to exploring high-performance
(Al,Cr)N DMS samples and understanding the high temperature
ferromagnetism.

The remaining part of this paper is organized as follows. We shall
present our computational method and parameters in next section.
In the third section, we shall present our main results of
optimized magnetic structures (through geometric and internal
atomic position optimization) and defect formation energies. In
the fourth section, we shall present main electronic structures,
including energy bands and electron density distributions. In the
fifth section, we shall discuss the ferromagnetism correlated with
N defects in real samples. Finally, we shall give our main
conclusion in the sixth section.

\section{Computational method and parameters}

\begin{figure}[tbh]
\scalebox{0.9}{\includegraphics[width=8cm]{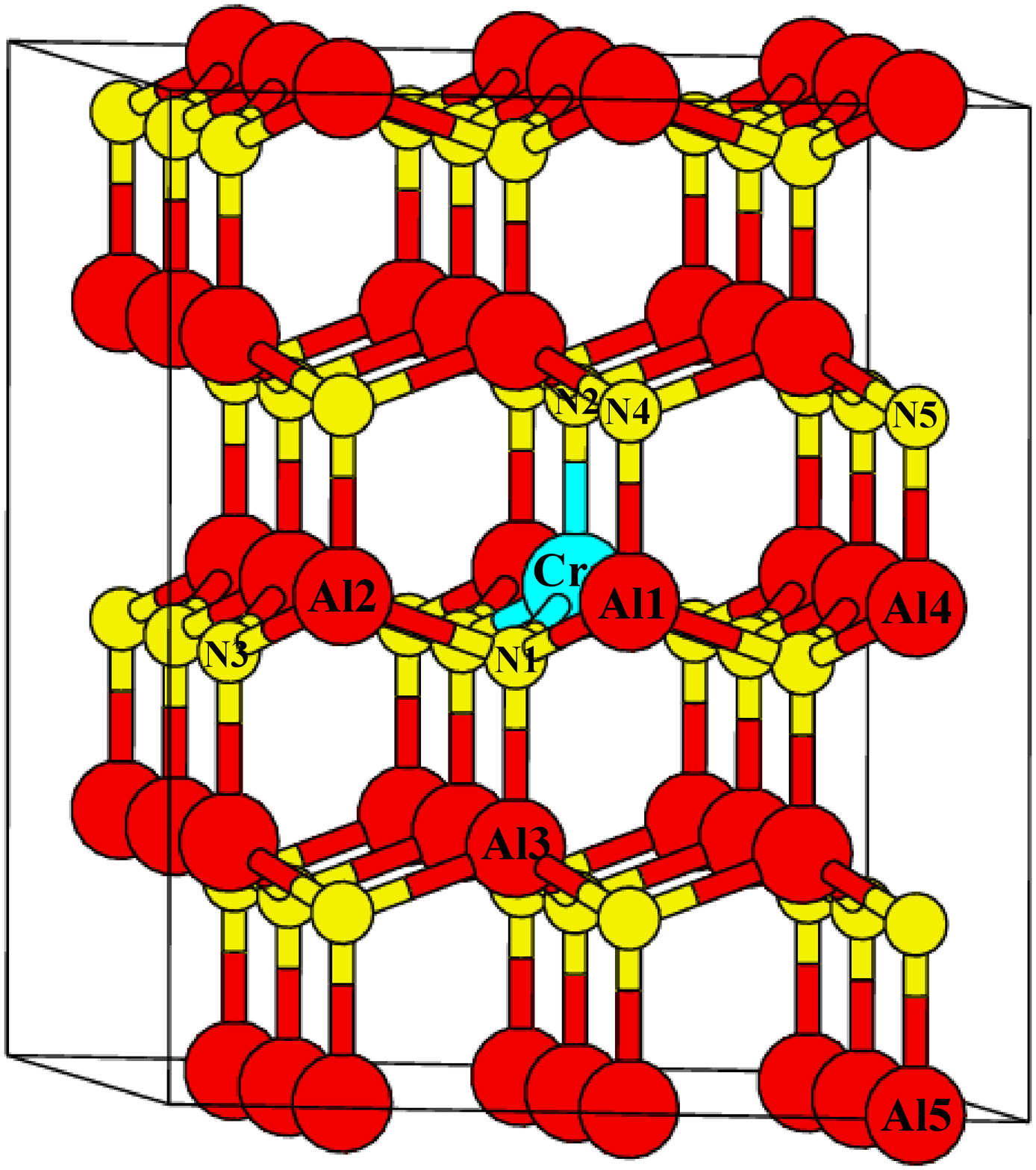}}
\caption{(color online). Schematic structure of Cr-substituted
3$\times$3$\times$2 wurtzite AlN supercell, CrAl$_{35}$N$_{36}$.
The centering largest (cyan or white) ball is the first Cr atom,
the smallest (yellow or light gray) ones are N atoms, and the
other (red or gray) ones are Al atoms. An N vacancy can be created
by removing an N atom, and the second substitutional Cr atom can
be introduced by substituting a Cr for an Al atom. For examples,
$V_{\rm N1}$ (or $V_{\rm 1}$) is made by removing N1, and Cr3 by
substituting a Cr atom for Al3.}
\end{figure}

We use a first-principles method and supercell approach to study
wurtzite AlN with doped Cr atoms and N defects (including nitrogen
vacancies and nitrogen interstitials). We use a
3$\times$3$\times$2 supercell as showed in Fig. 1 and substitute a
Cr atom for an Al atom to model Cr doping concentration 2.78$\%$
realized in MBE samples\cite{aln,aln900b,aln900c,aln-insua}. This
supercell is not very large to simulate real experimental
conditions, but it is large enough to model the representative
concentrations of both doped Cr and $V_{\rm N}$ and main
properties of Cr-Cr and Cr-$V_{\rm N}$ interactions, especially
considering that our calculations are done with full-potential
linear augmented plane wave (FLAPW) method within the
density-functional theory (DFT)\cite{DFT}. Using such supercell is
also supported by the following calculated results that the total
energy of both Cr$_{\rm Al}$-Cr$_{\rm Al}$ and Cr$_{\rm
Al}$-$V_{\rm N}$ pairs substantially increases with increasing the
Cr$_{\rm Al}$-Cr$_{\rm Al}$ and Cr$_{\rm Al}$-$V_{\rm N}$
distances, respectively. The supercell consists of one Cr atom, 35
Al atoms, and 36 N atoms and its space group is No. 156. Such a
supercell for the pure AlN consists of 36 Al and 36 N atoms. The
second Cr atom is introduced by substituting for one of Al atoms
in the supercell. For examples, Cr1 is made by substituting a Cr
atom for the Al atom labelled as Al1 in Fig. 1. A nitrogen vacancy
$V_{\rm N}$ is created by removing one N atom. For examples,
$V_{\rm N1}$ (or $V1$) is made by removing the N atom labelled as
N1. An interstitial N atom is introduced by adding an N atom in
the supercell. We obtain an octahedral N interstitial N$_{\rm T}$
if adding an N atom at an octahedral site, or obtain tetrahedral N
interstitial N$_{\rm T}$ if adding an N atom at a tetrahedral
site. We optimize lattice constants to minimize the total energy,
and at the same time optimize internal atomic coordinates to make
the force on every atom less than 0.04 eV/\AA. All the electronic
structures and magnetic moments are calculated with equilibrium
lattice constants.

All the presented results are calculated with the full-potential
linear augmented plane wave method within the density-functional
theory\cite{DFT}, as implemented in the package
WIEN2k\cite{wien2k}, although we also use pseudo potential method
to do some preliminary structural optimizations. The generalized
gradient approximation (GGA)\cite{PBE96} is used as the
exchange-correlation potential for all our presented results, and
some important cases are confirmed by local-density-approximate
calculations\cite{lsda}. Full relativistic calculations are done
for core states and the scalar approximation is used for the
others, with the spin-orbit coupling being neglected\cite{soc}.
The radii of muffin-tin spheres, $R_{\rm mt}$, are adjusted so as
to achieve the best accuracy of self-consistent calculations. We
use the parameters $R_{\rm mt}$$\cdot$$K_{\rm max}$=8.0 and make
the sphere harmonic expansion upto $l_{\rm max}$=10 in the
spheres. The Brillouin zone integrations are performed with the
special k-point method over a 5$\times$5$\times$3 Monkhorst-Pack
mesh. A convergence standard, $\int|\rho_n-\rho_{n-1}|dr <0.00005$
($\rho_n$ and $\rho_{n-1}$ are output and input charge density,
respectively), is used for all our self-consistent calculations.

\section{optimized structures and defect formation energies}

We optimize fully the lattice structures and internal atomic
positions for N vacancy (Al$_{36}$N$_{36}+V_{\rm
N}$=Al$_{36}$N$_{35}$), Cr doped AlN (CrAl$_{35}$N$_{36}$),
Cr-atom plus N-vacancy (CrAl$_{35}$N$_{36}+V_{\rm
N}$=CrAl$_{35}$N$_{35}$), Cr-atom plus tetrahedral N interstitial
 (CrAl$_{35}$N$_{36}+\rm N_{T}$), and Cr-atom plus octahedral
N interstitial  (CrAl$_{35}$N$_{36}+\rm N_{O}$). We can have
various atomic configurations for pairs of Cr-$V_{\rm N}$,
Cr-N$_{\rm T}$, and Cr-N$_{\rm O}$ with given distance, and they
correspond to different total energies. As for
CrAl$_{35}$N$_{36}+V_{\rm N}$, the total energy changes with the
distance between Cr and $V_{\rm N}$, and the lowest total energy
is reached when Cr and $V_{\rm N}$ are the nearest neighbors
Cr-$V1$ (connected with an approximately horizontal Cr-N bond).
The next lowest total energy (0.23 eV) is reached by Cr-$V2$. For
double Cr doped system, the total energy depends on the Cr-Cr
distance, and the lowest total energy is reached when the two Cr
(Cr and Cr1) are in the same layer and connected with the same N
(consistent with previous result\cite{alncr-dft}). The next lowest
total energy (0.28 eV) is reached by Cr-Cr3 and other equivalent
Cr-Cr pairs. These tends are shown in Fig. 2.

\begin{figure}[!tbh]
\scalebox{1}{\includegraphics[width=8.5cm]{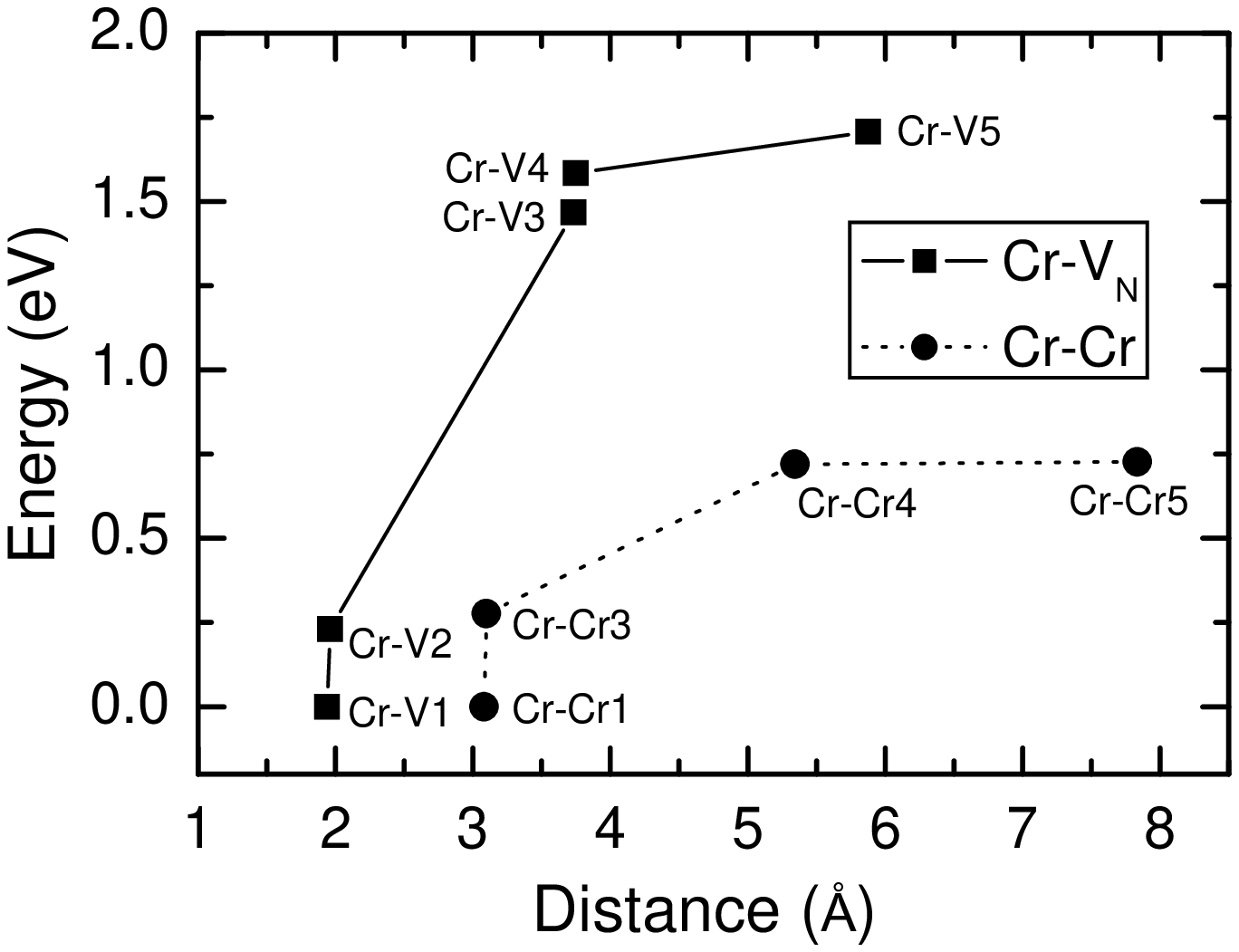}}
\caption{Relative total energy of the Cr doped AlN with N
vacancies for different Cr-$V_{\rm N}$ distances (filled squares)
and that of Cr doped AlN for different Cr-Cr distance (filled
circles). Here Cr-Cr$n$ means the pair of the first Cr and the
second Cr at the Al$n$ site, and Cr-V$n$ the pair of the first Cr
and the N vacancy at the N$n$ site, as are labelled in Fig. 1. The
two smallest total energies, corresponding to the nearest
Cr-$V_{\rm N}$ and Cr-Cr pairs, are set to zero.}
\end{figure}

\begin{table}[!tbh]
\caption{The calculated equilibrium lattice constants ($a$/$c$),
the relative volume change  ($\delta v$) with respect to the pure
AlN, the magnetic moment per magnetic atom ($M$), and the
electronic density of states at the Fermi level ($D_{\rm F}$). The
pure AlN phase is presented for comparison. \label{tab:1}}
\begin{ruledtabular}
\begin{tabular}{cccccc}
Name                  & $a$/$c$ (\AA) &$\delta v$ (\%)&  $M$ ($\mu_B$)& $D_{\rm F}$\\
\hline Al$_{36}$N$_{36}$+$V_{\rm N}$        &     9.65/10.29   &  1.74     &  1.000    & $>$0  \\
\hline CrAl$_{35}$N$_{36}$             &  9.59/10.26  &   0.21  &    3.000      &   $>$0 \\
\hline CrAl$_{35}$N$_{36}$+$V_{\rm N1}$             &  9.58/10.22  &  -0.16  &    4.000      &    0   \\
\hline CrAl$_{35}$N$_{36}$+$\rm N_{T1}$ &  9.66/10.26  &   1.28  &    2.000      &    0   \\
\hline CrAl$_{35}$N$_{36}$+$\rm N_{O1}$ &  9.61/10.29  &   1.15  &    4.000      &    0   \\
\hline Al$_{36}$N$_{36}$               &  9.58/10.27  &   0.00  &    0.000      &    0   \\
\end{tabular}
\end{ruledtabular}
\end{table}

Our systematic calculations show that a doped Cr atom contributes
3$\mu_B$ to the total magnetic moment, and a neighboring N-vacancy
adds 1$\mu_B$ because its electron just below the Fermi level is
transferred to the Cr $d$ state. Actually, an isolate N-vacancy,
even without any Cr doping, has a moment of 1$\mu_B$, but N
vacancies are in paramagnetic order because inter-vacancy spin
interaction is equivalent to zero within computational error. As
for interstitial N atoms, a tetrahedral N carries a moment of
-1$\mu_B$ and an octahedral N has 1$\mu_B$, but they are too high
in total energy to exist in real samples. Two neighboring Cr
atoms, such as Cr-Cr1 and Cr-Cr3, have 6$\mu_B$ because their
ferromagnetic couplings are favorable in energy. In fact, the
ferromagnetic orientation is lower in total energy by 464 meV and
195 meV than the corresponding antiferromagnetic one for Cr-Cr1
and Cr-Cr3, respectively. The energy difference 464 meV is
decreased to 163 meV when one N vacancy is at the nearest site,
and is further reduced to 25 meV when two N vacancies are in the
nearest neighborhood. This trend is consistent with effect of N
vacancy on Mn-Mn spin exchange constants\cite{gmn-dft3}. The
lattice constants, the relative volume changes, the magnetic
moments per supercell of the most stable configurations are
summarized in Table I. The most stable is the  N interstitial $\rm
N_{T1}$ ($\rm N_{O1}$), the nearest to the Cr atom, in the
CrAl$_{35}$N$_{36}$+$\rm N_{T1}$ (CrAl$_{35}$N$_{36}$+$\rm
N_{O1}$), among all the tetrahedral (octahedral) N interstitials.
The corresponding results for the pure AlN are also presented for
comparison. The DOS at the Fermi level, $D_F$, is zero in all the
cases except for the CrAl$_{35}$N$_{36}$ whose Fermi level is at
the Cr $d$ impurity bands and the Al$_{36}$N$_{36}+V_{\rm N}$
whose Fermi level is at the N-vacancy-induced in-gap band.

We further calculate formation energy for all the cases. The
formation energy of the supercell including $n_{V_{\rm N}}$ N
vacancies and $n_{\rm Cr}$ substitutional Cr atoms is defined
as\cite{zhang1,zhang2}
\begin{equation}\begin{array}{rl}
E_f(n_{V_{\rm N}},n_{\rm Cr}) =& \displaystyle E(n_{V_{\rm
N}},n_{\rm Cr})-E(0,0)\\[0.2mm] &\displaystyle  +n_{V_{\rm N}}\mu_{\rm
N}-n_{\rm Cr}\mu_{\rm Cr}+n_{\rm Cr}\mu_{\rm Al}\end{array}
\end{equation}
where $E(n_{V_{\rm N}},n_{\rm Cr})$ is the total energy of the
supercell, and ($\mu_{\rm N},\mu_{\rm Cr},\mu_{\rm Al}$) are
chemical potentials of N, Cr, and Al. Using $\mu^{\rm gas}_{\rm
N}$, $\mu^{\rm bulk}_{\rm Al}$, and $\mu^{\rm bulk}_{\rm Cr}$ as
the chemical potentials of N gas, Cr bulk, and Al bulk, we have
$\mu_{\rm N}=\mu^{\rm gas}_{\rm N}+\bar{\mu}_{\rm N}$, $\mu_{\rm
Al}=\mu^{\rm bulk}_{\rm Al}+\bar{\mu}_{\rm Al}$, and $\mu_{\rm
Cr}=\mu^{\rm bulk}_{\rm Cr}+\bar{\mu}_{\rm Cr}$. The
thermodynamical stability of AlN requires
\begin{equation}\label{aln}
\bar{\mu}_{\rm N}+\bar{\mu}_{\rm Al} = \bar{\mu}_{\rm AlN}.
\end{equation}
The fact that there is no spontaneous formation of CrN implies
\begin{equation}\label{crn}
\bar{\mu}_{\rm N}+\bar{\mu}_{\rm Cr} \leq \bar{\mu}_{\rm CrN}.
\end{equation}
We can express the minimal formation energy as
\begin{equation}\begin{array}{c}
\displaystyle E^{\rm min}_f(n_{V_{\rm N}},n_{\rm Cr}) =
E(n_{V_{\rm N}},n_{\rm Cr})-E(0,0)\\[0.4mm] \displaystyle  +n_{V_{\rm
N}}\mu^{\rm gas}_{\rm N}-n_{\rm Cr}(\mu^{\rm bulk}_{\rm
Cr}-\mu^{\rm bulk}_{\rm
Al})\\[0.4mm]
\displaystyle  +n_{V_{\rm N}}\bar{\mu}^{\rm min}_{\rm N}-n_{\rm
Cr}(\bar{\mu}^{\rm max}_{\rm Cr}-\bar{\mu}^{\rm min}_{\rm
Al})\end{array}\label{forme}
\end{equation}
where the superscripts max (min) means the maximum (minimum) of
the chemical potential. The chemical potentials $\mu^{\rm
gas}_{\rm N}$, $\mu^{\rm bulk}_{\rm Al}$, and $\mu^{\rm bulk}_{\rm
Cr}$ can be calculated in terms of total energies of N gas, Al
bulk, and Cr bulk. The maximal and minimal chemical potentials can
be obtained in terms of equality (\ref{aln}) and inequality
(\ref{crn}) and, therefore, the minimal formation energy can be
calculated in terms of Eq. (\ref{forme}) as long as
$\bar{\mu}_{\rm CrN}$ and $\bar{\mu}_{\rm AlN}$ are known.

\begin{table}[!thb]
\caption{The formation energies ($\Delta E$) of a substitutional
Cr atom and an N vacancy, and their pairs in AlN. The `+', `0',
and `-' superscripts mean that the extra charge is equivalent to
$e$, 0, and -$e$, where $e$ is the absolute value of the
electronic charge. The last row corresponds to the case that the
distance between Cr and $V_{\rm N}$ is the longest within the
3$\times$3$\times$2 supercell, and the fourth row the case that
the distance is the shortest.}
\begin{ruledtabular}
\begin{tabular}{cccc}
& Name         & $\Delta E$ (eV) & \\
\hline & Al$_{36}$N$_{36}$$+V_{\rm N1}$     &      -0.23 & \\
\hline & Al$_{36}$N$_{36}+$$V_{\rm N}^+$ &      0.1 \cite{aln-form} & \\
\hline & Cr$^0$Al$_{35}$N$_{36}$         &      0.04 & \\
\hline & Cr$^-$Al$_{35}$N$_{36}$+$V_{\rm N1}^+$  &      0.77 & \\
\hline & Cr$^-$Al$_{35}$N$_{36}$+$V_{\rm N5}^+$  &      2.47 & \\
\end{tabular}
\end{ruledtabular}
\end{table}

Our calculated result is $\bar{\mu}_{\rm AlN}=-3.33$ eV for AlN,
and we use known data $\bar{\mu}_{\rm CrN}=-1.30$ eV for
CrN\cite{jap83}. For a single neutral N-vacancy in AlN under poor
N condition, we have $n_{V_{\rm N}}=1$ and $n_{\rm Cr}=0$, and
thus the formation energy is -0.23 eV, while the formation energy
of a single N$^+$ vacancy is 0.1 eV\cite{aln-form}. For a neutral
Cr atom under rich N condition, we have $n_{V_{\rm N}}=0$ and
$n_{\rm Cr}=1$, and thus the formation energy is 0.04 eV. This
small formation energy 0.04 eV is reasonable and consistent with
experimental high Cr doping
concentrations\cite{aln,aln900a,alncr-vrh,aln-npres,aln-otherb}. A
previous unreasonably high formation energy 3.01
eV\cite{aln-cr-cluster} is in contradiction with the experimental
high Cr doping concentrations. Charged Cr atoms have much higher
formation energies. For the nearest Cr-$V_{\rm N1}$ pair, we have
$n_{V_{\rm N}}=1$ and $n_{\rm Cr}=1$, and thus the formation
energy is 0.77 eV, while more separated Cr-$V_{\rm N5}$ pairs have
higher formation energies. These results are summarized in Table
II. Interstitial N atoms, both tetrahedral and octahedral, have
very high total energies and formation energies, and therefore are
not presented in Table II. This implies that either dilute N
vacancies or dilute Cr atoms are easy to be realized in AlN
because the interactions can be neglected in the dilute doping
limit.

\section{electronic structures}

\begin{figure}[tbh]
\includegraphics[width=8.5cm]{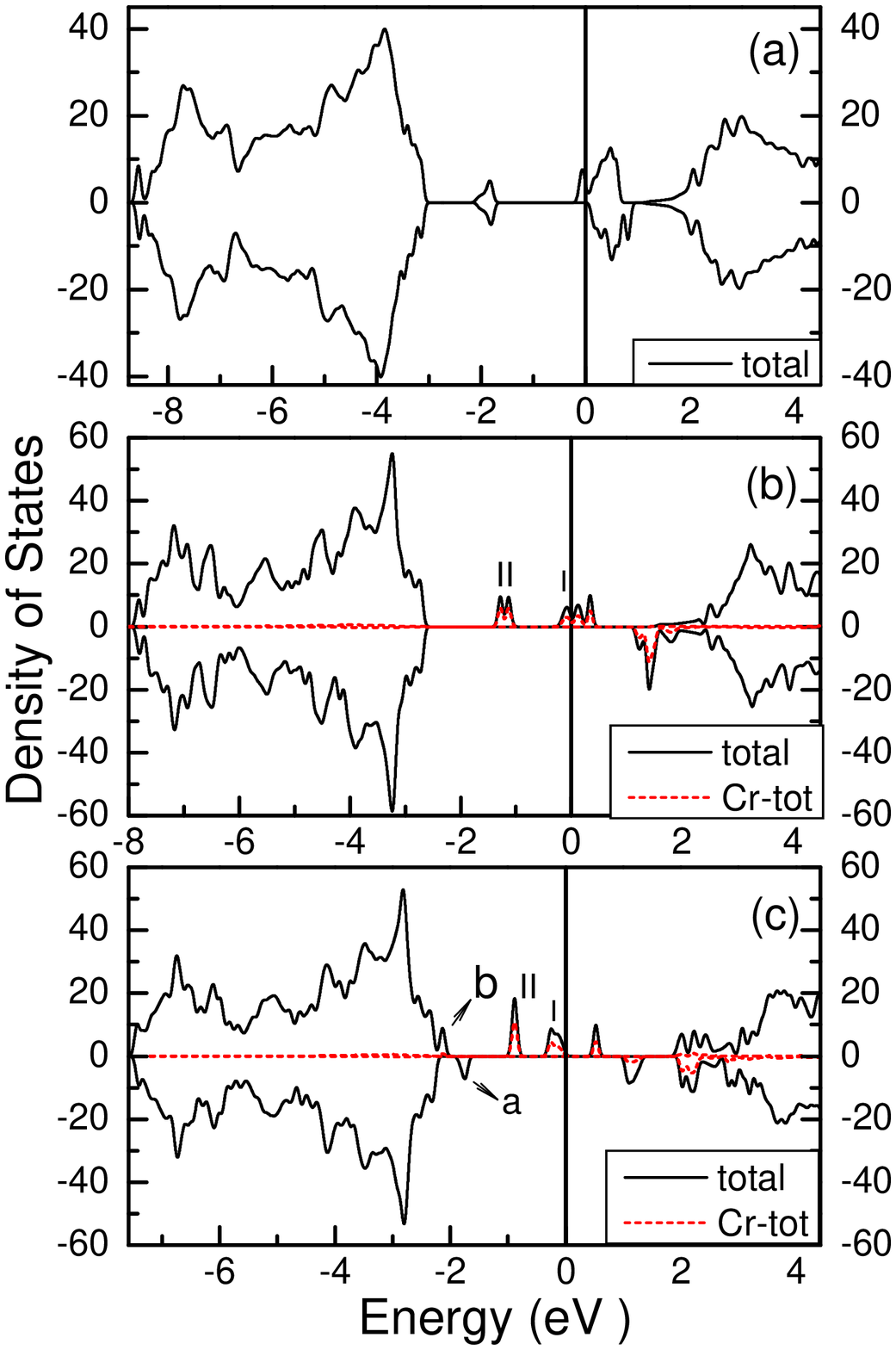}
\caption{(color online). Spin-dependent total DOS (black solid
lines) and Cr $d$ partial DOS (red or gray dash lines) of the
Al$_{36}$N$_{35}$ (including one $V_{\rm N}$) (a), the
CrAl$_{35}$N$_{36}$ (including one Cr atom) (b), and the
CrAl$_{35}$N$_{35}$ (including one Cr+$V_{\rm N1}$ complex) (c).
The upper half for each panel is for majority spin channel and the
other for minority spin. For the Al$_{36}$N$_{35}$, the filled
part just below the Fermi level contains on electron, and thus one
$V_{\rm N}$ contributes 1$\mu_B$. For the CrAl$_{35}$N$_{36}$,
there is one electron in the small part `I' and two in the filled
part of `II', and hence the total moment is 3$\mu_B$. For the
CrAl$_{35}$N$_{35}$, both `I' and `II' are filled with two
electrons, and the total moment is 4$\mu_B$. The small parts `a'
and `b', both filled with one electron, are due to $V_{\rm N}$.}
\end{figure}

\begin{figure}[tbh]
\includegraphics[width=8.5cm]{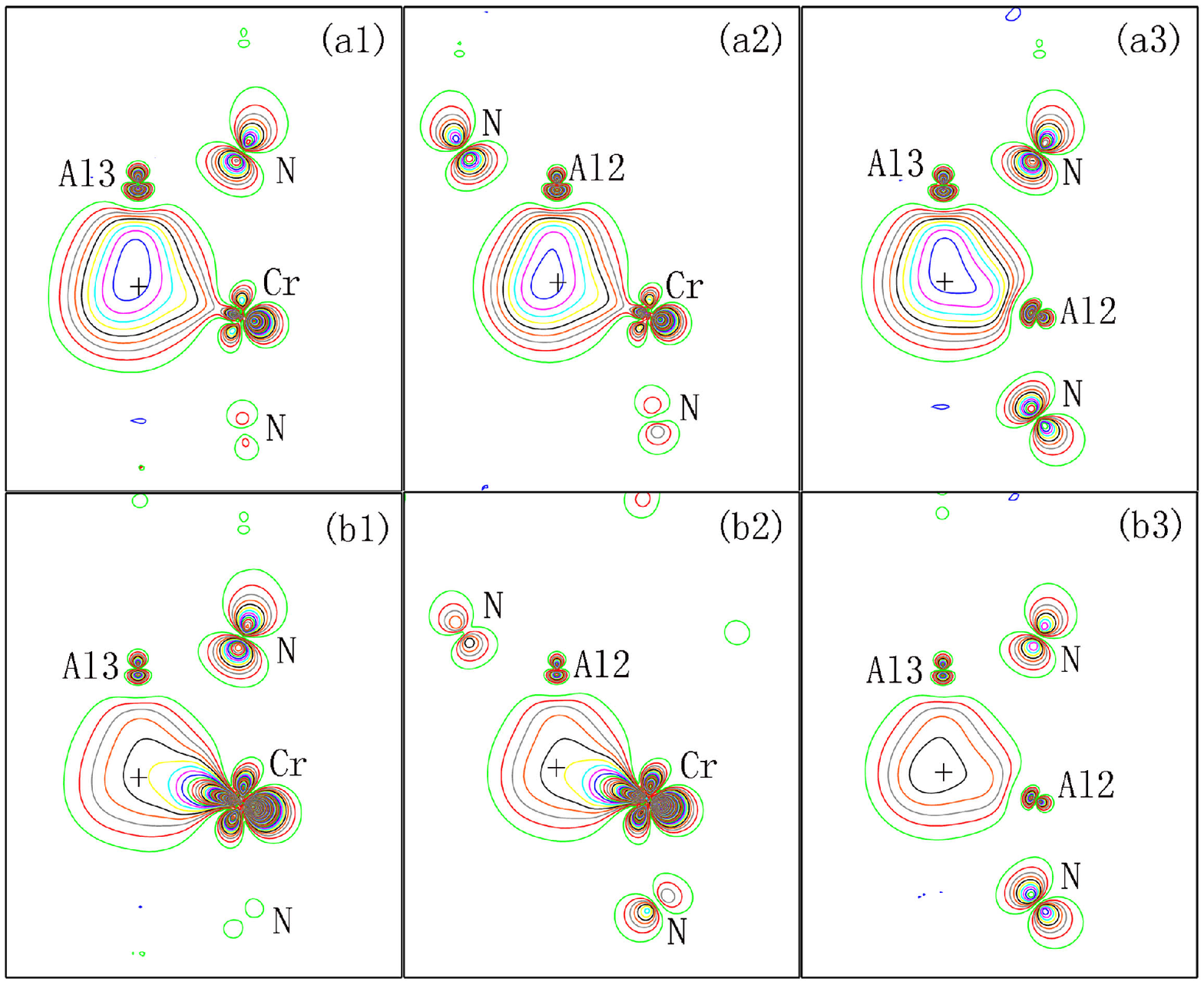}
\caption{(color online). Electron density distributions of the `a'
and `b' bands of the CrAl$_{35}$N$_{35}$, defined in panel (c) of
Fig. 3. The a1, a2, and a3 (b1, b2, and b3) show the electron
density distributions of the `a' (`b') bands in the three planes
defined by Cr-$V_{\rm N1}$-Al3, Cr-$V_{\rm N1}$-Al2, and
Al2-$V_{\rm N1}$-Al3, where Cr, Al1, Al2, Al3 means the atoms
labelled with the same signs in Fig. 1 and the $V_{\rm N1}$ is at
the N1 site, labelled as `+', connecting the Cr atom and three Al
ones. The outmost (green or light gray) line represents 0.001
$e$/a.u.$^3$. The density contour increment is 0.001
$e$/a.u.$^3$.}
\end{figure}

We present spin-dependent DOS of the Al$_{36}$N$_{35}$ (including
one N vacancy in the supercell), the CrAl$_{35}$N$_{36}$
(including one Cr substitutional at an Al site), and the
CrAl$_{35}$N$_{35}$ (including one Cr+$V_{\rm N1}$ complex) in
Fig. 3. The valence bands for every spin channel are filled with
108 electrons for the AlN supercell, and in total we have 216
electrons in the valence bands. The absence of one N atom removes
3 $p$ electrons and re-organizes the band structures, as shown in
Fig. 3(a). Now there are three electrons in the gap of AlN and the
valence band manifold is filled with 210 electrons, 105 for each
spin channel. The N vacancy cannot donate electron to the
conduction bands because the Fermi level is too low with respect
to the conduction bands. This is different from the N-vacancy of
GaN\cite{gan-nvac}. Fig. 3.(b) shows the DOS of the
CrAl$_{35}$N$_{36}$. In this case we have 108 electrons in the
valence band manifold for each spin channel and the Cr $d$ states
are in the gap of AlN. The $d$ states are split into a doublet
(labelled as `II') and a triplet (labelled as `I'). The former is
filled with two electrons and the latter by one. The Fermi level
is in the bands of the triplet. Hence one Cr contributes 3$\mu_B$
to the moment. In Fig. 3(c) we show the density of states of the
CrAl$_{35}$N$_{35}$. The interaction of the doped Cr and the N
vacancy makes the filled singlet originating from the N vacancy
move lower in energy and become two spin-split parts: `a' and `b',
and at the same time makes the Cr $d$ triplet split into a doublet
(also labelled as `I' in Fig. 3(c)) and a singlet. The valence
band manifold below the part `b' consists of 105+105 filled
states. As a result, one Cr+$V_{\rm N}$ complex contributes
4$\mu_B$ to the moment.

\begin{figure}[t]
\includegraphics[width=8.5cm]{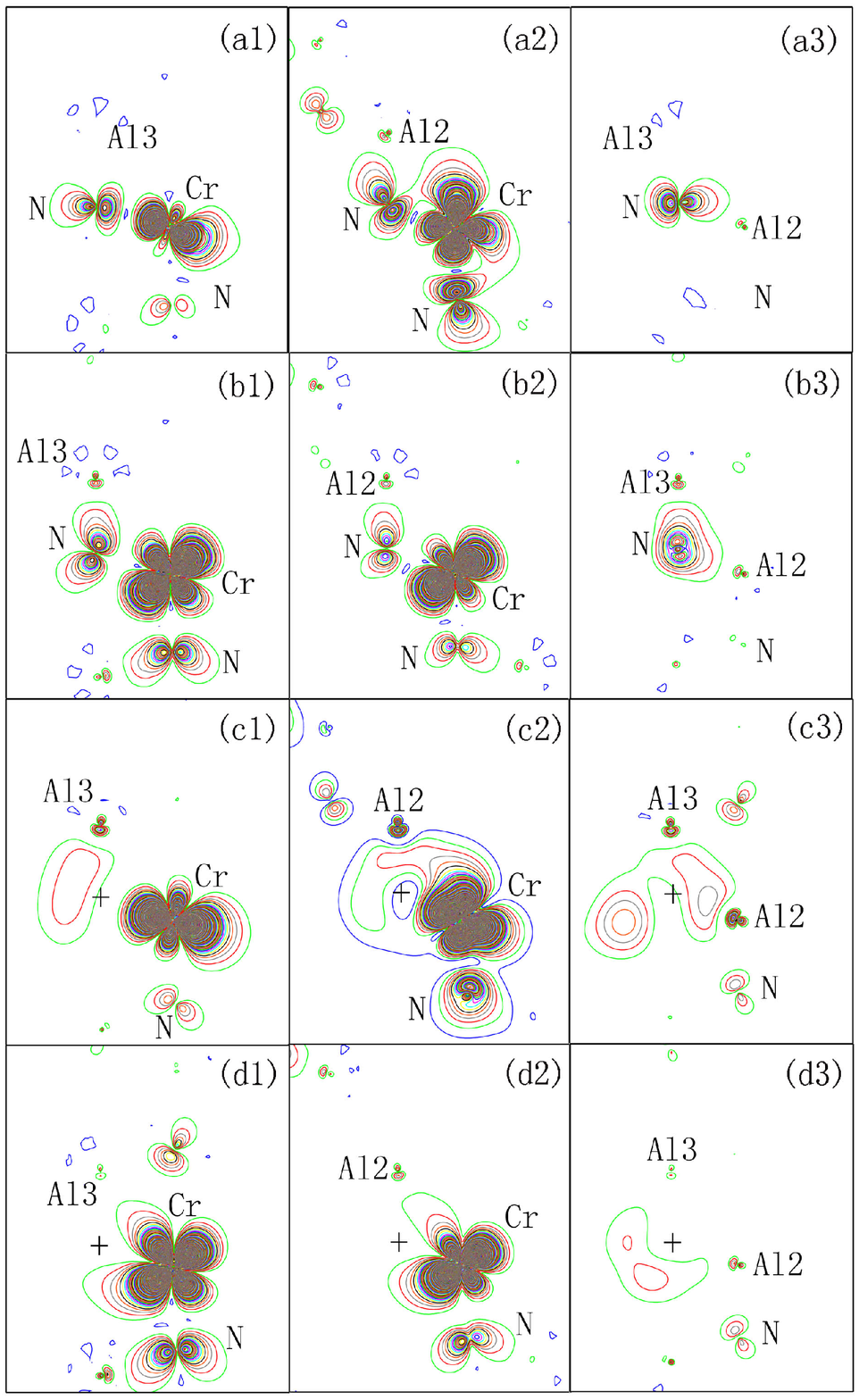} \caption{(color
online). Electron density distributions of the parts `I' (a1, a2,
a3) and `II' (b1, b2, b3) of the CrAl$_{35}$N$_{36}$, defined in
panel (b) of Fig. 3, and those of the parts `I' (c1, c2, c3) and
`II' (d1, d2, d3) of the CrAl$_{35}$N$_{35}$, defined in panel (c)
of Fig. 3, where the three planes are the same as those in Fig. 4.
It should be pointed out that the part `I' of the
CrAl$_{35}$N$_{35}$ consists of two electrons, but that of the
other consists of one electron only. The outmost (green or light
gray) line represents 0.001 $e$/a.u.$^3$. The density contour
increment is 0.001 $e$/a.u.$^3$. }
\end{figure}

We present in Fig. 4 electron density distribution of the
N-vacancy-induced filled states, the parts `a' and `b' indicated
in Fig. 3(c), of the CrAl$_{35}$N$_{35}$. The three planes are
defined by three-site series: Cr-$V_{\rm N}$-Al3, Cr-$V_{\rm
N}$-Al2, and Al2-$V_{\rm N}$-Al3. Here Cr, Al1, Al2, Al3 are the
atoms labelled with the same signs in Fig. 1. It can be seen that
the charge density distribution at the N vacancy is about
0.014$e$/a.u.$^3$, where $e$ is the absolute value of the
electronic charge. The total charge within the N vacancy is
estimated to be approximately equivalent to 1.1$e$. The remaining
0.9$e$ charge of the parts `a' and `b' is distributed at the
nearest Cr site (0.6$e$) and N site (0.3$e$). This indicates that
the N-vacancy state is substantially hybridized with the Cr $d$
states and substantial charge is transferred to the vacancy from
the neighboring Al (or Cr) atoms although there is no N atom at
the site.

We present in Fig. 5 the Cr $d$ electron density distribution of
the CrAl$_{35}$N$_{36}$ and the CrAl$_{35}$N$_{35}$. The three
planes are the same as those in Fig. 4. For the
CrAl$_{35}$N$_{36}$, it is estimated approximately that  80\% of
the `II' charge is around the Cr atom and 70\% of the `I' charge
around the Cr, and the Cr electronic configuration is $d^3$. For
the CrAl$_{35}$N$_{35}$, approximately 90\% of the `II' charge is
around the Cr atom and 85\% of the `I' charge around the Cr. The
remaining charge is around the nearest N sites. Therefore, we can
conclude that Cr electronic configuration in a Cr+$V_{\rm N}$
complex is $d^4$, not $d^3$. The extra electron comes from the N
vacancy. Thus the valent state of the Cr+$V_{\rm N}$ complex is
Cr$^-$+$V_{\rm N}^+$.

\section{ferromagnetism correlated with N defects in real samples}

Because the minimal formation energy for forming a neutral N
vacancy in AlN is -0.23 eV, a small concentration of N vacancies
can be formed spontaneously in AlN under poor N condition. An N
vacancy contributes 1$\mu_B$ to the moment, but our calculations
show that spin interaction between N-vacancy spins is estimated to
be zero within computational error. Hence N vacancies alone cannot
form ferromagnetism. Because the minimal formation energy for
forming a substitutional Cr in AlN is 0.04 eV, dilute Cr doping
alone is easy to be realized in AlN at room temperature under
N-rich condition. High concentrations of Cr atoms have been doped
into AlN\cite{aln,aln900a,alncr-vrh,aln-npres,aln-otherb},
although the Cr formation energy may be larger when Cr-Cr
interaction cannot be neglected. This implies that there is no
barrier of formation energy in the dilute doping limit. The Cr
doped AlN can be ferromagnetic through some mechanism based on
hybridization between local Cr-$d$ states and extended N-$p$
states\cite{gmn-dft1,gmn-dft2,alncr-slj}.

The minimal formation energy for forming a Cr+$V_{\rm N}$ complex
is 0.77 eV. As a result, the concentration of such complex is at
most 2.3$\times$10$^{-6}$ per Al site when the synthesis
temperature is at 600 K, becomes 10$^{-3}$ at 1140 K, and reaches
to 1\% at 1700 K. Each of the Cr+$V_{\rm N}$ complexes contributes
4$\mu_B$ to the total moment. It should be pointed out that the
maximal equilibrium concentration of such complexes at 1000 K is
lower at least by two orders of magnitude than the Cr
concentrations  of MBE-grown samples (nominally $2\sim
7$\%)\cite{aln900b,aln900c,aln-insua,aln-insub} and other ones
(nominally $3\sim
35$\%)\cite{aln,aln900a,alncr-vrh,aln-npres,aln-otherb} of
(Al,Cr)N diluted magnetic semiconductors. Thus the synthesis of
real samples is not through growth of the Cr+$V_{\rm N}$
complexes.

\begin{figure}[tbh]
\includegraphics[width=7.5cm]{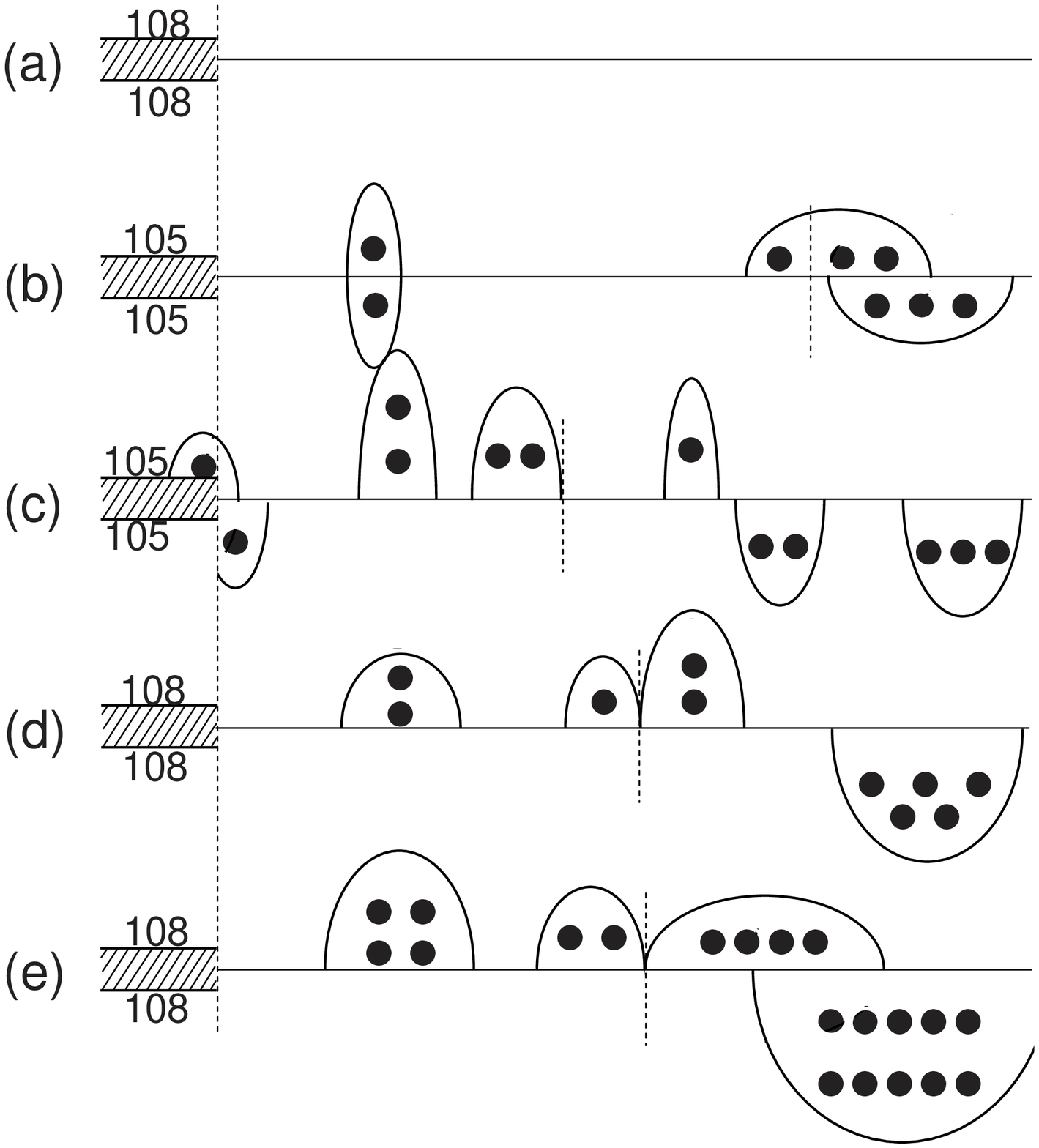} \caption{ Schematic
DOS of the Al$_{36}$N$_{36}$ (a), the Al$_{36}$N$_{35}$ (including
one $V_{\rm N}$) (b), the CrAl$_{35}$N$_{35}$ (including one
Cr+$V_{\rm N1}$ complex) (c), the CrAl$_{35}$N$_{36}$ (including
one substitutional Cr atom) (d), and the Cr$_2$Al$_{34}$N$_{36}$
(including two nearest substitutional Cr atoms) (e). The two
numbers for the shadow region denote the numbers of the
majority-spin $p$ valence bands and minority-spin ones. A solid
dot means a filled electron in the bands.}
\end{figure}

It is reasonable to suppose that there coexist N vacancies, Cr
atoms, and Cr+$V_{\rm N}$ complexes in real samples and the
fractional concentrations are dependent on synthesis
conditions\cite{aln900b,aln900c}. Because of N pressure
fluctuation during synthesis, a substitutional Cr and an N vacancy
can be formed easily in different regions. This implies that some
N-vacancy domains and some Cr-doping domains can be formed
simultaneously in separated different regions. The former are
paramagnetic and the latter are ferromagnetic, but these domains
are subject to thermodynamical equilibrium of electrons in real
samples. We present the schematic DOS and the corresponding Fermi
levels of the Al$_{36}$N$_{36}$, the Al$_{36}$N$_{35}$ (including
one N vacancy in the supercell), the CrAl$_{35}$N$_{35}$
(including one Cr+$V_{\rm N1}$ complex), the CrAl$_{35}$N$_{36}$
(including one Cr substitutional at an Al site), and the
Cr$_2$Al$_{34}$N$_{36}$ (including Cr and Cr1 as defined in Fig.
1) in Fig. 6 (a $\sim$ e). The Fermi level of the
CrAl$_{35}$N$_{36}$ is consistent with that of the
Cr$_2$Al$_{34}$N$_{36}$. Because the Fermi level of the N vacancy
is substantially higher than that of the Cr atom, the electron of
the N vacancy must transfer to the Cr site and this electron
transfer results in pairs of $V^+_{\rm N}$ and Cr$^-$. The Coulomb
attraction between $V^+_{\rm N}$ and Cr$^-$ will make them tend to
move together and form the neutral Cr+$V_{\rm N}$ complex. Both of
the Fermi levels of the $V^+_{\rm N}$ and the neutral Cr+$V_{\rm
N}$ complex are consistent with that of Cr atom. The electron of
the N vacancy tends to be transferred to the Cr atom and the
neutral Cr+$V_{\rm N}$ complex.

The N vacancy is in favor of the ferromagnetism through forming
the Cr+$V_{\rm N}$ complex which contributes 4$\mu_B$ to the
moment and Cr+2$V_{\rm N}$ complex which contributes 5$\mu_B$. On
the other hand, the N vacancy can be harmful to the ferromagnetism
through destroying the ferromagnetic spin coupling between two
nearest substitutional Cr atoms. We can figure out what will
happen in terms of the two characteristic synthesis temperatures,
$T_{s1}$ and $T_{s2}$ ($>T_{s1}$), for
Cr$_{x}$Al$_{1-x}$N$_{1-y}$, where $x$ is Cr concentration and $y$
is N-vacancy concentration. We suppose that there is no second
phase and hence set $y<2x$ in the following\cite{aln-secod}. When
$T<T_{s1}$, neither N-vacancy nor Cr atom can diffuse. A little of
the N-vacancy electrons will transfer to Cr domains but further
electron transferring will be blocked by the built-in electric
field formed by the transferred electrons. As a result, the main
feathers of the domain patterns will remain. The Cr moments are
coupled by thermally-activated variable-range hopping of
electrons\cite{alncr-vrh}, and thus the N-vacancy enhances the
ferromagnetism\cite{aln-insub,aln-npres}. When $T_{s1}<T<T_{s2}$,
N-vacancy can diffuse, but Cr atom cannot. All the N-vacancies and
their electrons will transfer to Cr domains, and many Cr+$V_{\rm
N}$ and Cr+2$V_{\rm N}$ complexes are formed. In this case, the
coupling of moments is the same, but the conductivity becomes
weaker because an additional barrier must be overcome for the
electron to hop\cite{alncr-vrh}, and thus the enhancement
decreases with increasing $y$. When $T>T_{s2}$, both N-vacancy and
Cr atom can diffuse. Cr atoms tend to form
clusters\cite{aln-cr-cluster} (similar to Cr- and Mn-doped
GaN\cite{gan-cr-cluster1,gan-cr-cluster2,gan-cr-cluster3}) and the
N vacancies tend to change the inter-Cr ferromagnetic coupling
into antiferromagnetic coupling. In this temperature range, the
N-vacancy is harmful to the ferromagnetism. However, The N vacancy
can be made to enhance the ferromagnetism by keeping the synthesis
temperature below $T_{s2}$ so that Cr clusters are unlikely. This
is in good agreement with experimental observation that N
vacancies enhance the ferromagnetism in samples made at low
temperature\cite{aln-insub,aln-npres}.

\section{Conclusion}

We have investigated nitrogen defects and their effects on the
ferromagnetism of Cr-doped AlN with N vacancies using the
full-potential density-functional-theory method and supercell
approach. We investigate the structural and electronic properties
of the $V_{\rm N}$, single Cr atom, Cr-Cr atom pairs, Cr-$V_{\rm
N}$ pairs, and so on. In each case, the most stable structure is
obtained by comparing different atomic configurations optimized in
terms of the total energy and the force on every atom, and then is
used to calculate the defect formation energy and study the
electronic structures. Without N vacancy, two Cr atoms have the
lowest energy when they stay at the nearest Al sites and their
moments orient in parallel. An N vacancy and a Cr atom tend to
form a stable complex of Cr$^{-}$ and $V_{\rm N}^{+}$ and the
moment of 4$\mu_B$ origins from the Cr $d^4$ electrons. Our
calculated formation energies indicate that $V_{\rm N}$ regions
can be formed spontaneously under N-poor condition or high enough
concentrations of Cr atoms can be formed under N-rich condition,
but only tiny concentrations of Cr+$V_{\rm N}$ complexes can be
realized under usual experimental conditions. Both of the N
vacancy and Cr atom create states in the semiconductor gap of AlN
and the highest filled state of the N vacancy is higher than the
filled Cr states. The $V_{\rm N}$ enhances the ferromagnetism by
giving one electron to Cr ion to add 1$\mu_B$ to the Cr moment,
but reduces the ferromagnetic exchange constants between the spins
in the nearest Cr-Cr pairs.

The formation energy results imply that real Cr-doped samples with
N vacancies are formed by forming Cr-doped regions and some
separated $V_{\rm N}$ regions and through subsequent atomic
relaxation during annealing. The $V_{\rm N}$ enhancement of the
ferromagnetism is in good agreement with experimental observation
that low nitrogen pressure enhances the ferromagnetism of
Al$_{1-x}$Cr$_x$N$_{1-y}$ when the synthesis temperature is low so
that Cr atoms cannot diffuse\cite{aln-insub,aln-npres}. On the
other hand, too many N vacancies are harmful to the ferromagnetism
when the synthesis temperature is high enough to make Cr atoms
diffuse and tend to become clusters of secondary phases in which
Cr-Cr spin interactions mainly are
antiferromagnetic\cite{aln-insua,aln-insub,aln-othera,aln-otherb,aln-npres}.
However, we can make N-vacancies enhance the ferromagnetism by
keeping the synthesis temperature low enough to avoid Cr
diffusion. All these first-principles results are useful to
exploring high-performance Cr-doped AlN DMS materials and
understanding their high temperature ferromagnetism.

\begin{acknowledgments}
This work is supported  by Nature Science Foundation of China
(Grant Nos. 10874232, 10774180, 90406010, and 60621091), by the
Chinese Academy of Sciences (Grant No. KJCX2.YW.W09-5), and by
Chinese Department of Science and Technology (Grant No.
2005CB623602).
\end{acknowledgments}

\end{document}